\documentclass[aps,prd,preprintnumbers,nofootinbib,floatfix,11pt]{revtex4}

\usepackage{amsfonts,amsmath,amssymb,mathrsfs,graphicx,epsfig,color,ulem} 
\usepackage{hyperref}
\allowdisplaybreaks[1]
\newcommand{\be}{\begin{equation}}
\newcommand{\ee}{\end{equation}}
\newcommand{\ba}{\begin{eqnarray}}
\newcommand{\ea}{\end{eqnarray}}

\newcommand{\Cr}[1]{{{\cal O} \left( r^{#1} \right)}}

\newcommand{\red}[1]{{{{#1}}}}

\begin{document}
\begin{flushright}
    YITP-21-54, KOBE-COSMO-21-11, OCU-PHYS-541, AP-GR-169
    \end{flushright}

\title{Asymptotic behavior of null geodesics near future null infinity: Significance of gravitational waves}
\author{Masaya Amo$^{1}$, Keisuke Izumi$^{2,3}$, Yoshimune Tomikawa$^{4}$, Hirotaka Yoshino$^{5,6}$, and Tetsuya Shiromizu$^{3,2}$}

\affiliation{$^{1}$Center for Gravitational Physics, Yukawa Institute for Theoretical Physics, Kyoto University, 606-8502, Kyoto, Japan}
\affiliation{$^{2}$Kobayashi-Maskawa Institute, Nagoya University, Nagoya 464-8602, Japan} 
\affiliation{$^{3}$Department of Mathematics, Nagoya University, Nagoya 464-8602, Japan}
\affiliation{$^{4}$Faculty of Economics, Matsuyama University, Matsuyama 790-8578, Japan}
\affiliation{$^{5}$Advanced Mathematical Institute, Osaka City University, Osaka 558-8585, Japan}
\affiliation{$^{6}$Department of Physics, Kobe University, Kobe 657-8501, Japan}

\begin{abstract}
\begin{center}
{\bf Abstract}
\end{center}
\noindent
We investigate the behavior of null geodesics
near future null infinity in asymptotically flat spacetimes.
In particular, 
we focus on the asymptotic behavior of null geodesics
that correspond to worldlines of photons
initially emitted in the directions tangential to the
constant radial surfaces in the Bondi coordinates.
The analysis is performed for general dimensions,
and the difference between the four-dimensional cases and
the higher-dimensional cases is stressed.
In four dimensions, some assumptions are required
to guarantee the null geodesics to reach future null infinity,
in addition 
to the conditions of asymptotic flatness.
Without these assumptions, gravitational waves 
may prevent photons from reaching null infinity.
In higher dimensions, by contrast,
such assumptions are not necessary, and 
gravitational waves do not affect the asymptotic behavior of null geodesics.
\end{abstract}

\maketitle

%
%
%======================================%
%<<<<<<<<<<<< SECTION I  >>>>>>>>>>>>>>%
%======================================%
%
\section{Introduction}

In the past few years, the LIGO and Virgo collaborations have reported
many detections of the gravitational wave events \cite{Abbott:2016blz,Abbott:2020gyp} and opened a new era of gravitational wave astronomy.
The Event Horizon Telescope Collaboration has recently observed the black hole shadow at the center of the galaxy M87 \cite{Akiyama:2019cqa}. 
The observational progress motivates us to examine the asymptotic behavior of null geodesics near future null infinity. Since the geometric structures 
in the neighborhood of infinity are close to those of Minkowski spacetime,
one may naively expect that it would be rather simple.
However, this is not the case. 
Indeed, in four dimensions, it is well known that the 
supertranslation, which is a part of the asymptotic symmetries
of null infinity~\cite{Bondi,Sachs}, gives an observational effect
through the effect of the gravitational wave memory \cite{Zeldovich:1974,Christodoulou:1991,Thorne:1992}.
It is pointed out that the signal of gravitational wave
memory could be detected statistically by accumulating data
of gravitational waves observed at
ground-based interferometers (see, e.g., Refs.~\cite{Lasky:2016,Hubner:2021}).
Furthermore, the space-based detector,
the Laser Interferometer Space Antenna (LISA),
which is planned to be launched in the 2030s \cite{Audley:2017drz},
may be able to 
detect it directly in the observation of
supermassive black hole mergers \cite{Favata:2009ii}. Moreover, the supertranslation  
has attracted much attention to solve the information loss paradox \cite{Hawking:1976ra}
in the evaporation of black holes
due to the Hawking radiation~\cite{Hawking:2016msc} .

In this paper, we examine the behavior of the null geodesics
that correspond to worldlines of photons 
emitted in the direction
tangent to the constant radial surfaces
in the neighborhood of future null infinity.  In four dimensions, 
we clarify the sufficient conditions that guarantee
null geodesics to reach future null infinity.
Those sufficient conditions exclude the possibility that  
gravitational waves may significantly affect the fate of emitted photons, 
and the relation to the supertranslation is discussed. 
To compare with the four-dimensional cases,
we will also address the higher-dimensional cases,
where the supertranslation is absent \cite{Tanabe:2011es,Hollands:2016oma}. 
It will be clarified that the asymptotic behavior of null geodesics is
not affected by gravitational waves in higher dimensions. 

In Minkowski spacetime, any photon 
emitted in a direction tangential to the constant radial surface
arrives at future null infinity
by increasing the value of the radial coordinate $r$ of its position
unboundedly while keeping the value of the null coordinate $u$ finite.
By contrast, in an environment with strong gravitational field,
the situation is different.  
In a Schwarzschild spacetime, for example,
there exist null geodesics that
are wholly included in the hypersurface at $r=3M$. 
The hypersurface at $r=3M$ is called the photon sphere \cite{Virbhadra:1999}
or the photon surface \cite{Claudel:2000},
and null geodesics on it extend toward future timelike infinity $i^+$,
not future null infinity $\mathscr{I}^+$. 
Moreover, all photons emitted to angular directions in $r<3M$
fall into the black hole
because of the strong gravitational attraction. 
If one restricts attention on asymptotic regions, one may
expect that 
the situation would be similar to that of Minkowski spacetime; 
{\it i.e.}, the value of the radial coordinate $r$ is
naively expected to increase for photons emitted
in angular directions. 
We study whether such naive expectation is correct or not
and clarify the fact that  
there is a possibility that gravitational waves pull the photon inside. 

The rest of this paper is organized as follows. In Sec.~\ref{Future Null Infinity}, we give a brief introduction of asymptotically flat spacetimes
in terms of the Bondi coordinates and present initial conditions
for the geodesic equations. In Sec.~\ref{4d}, we study
the asymptotic behavior of null geodesics that correspond
to photons emitted in angular directions near future null infinity of four-dimensional spacetimes.
In Sec.~\ref{highd}, we examine higher-dimensional cases.
Section~\ref{summary} is devoted to a summary and discussion.
In Appendix~\ref{appendixa}, we present the components of the Christoffel symbols in the Bondi coordinates.
In Appendix~\ref{appendixb}, we give some details of
our analysis presented in the main article.

%
%
%======================================%
%<<<<<<<<<<<< SECTION II  >>>>>>>>>>>>>>%
%======================================%
%
\section{Brief Review of Null Infinity and Initial Conditions}
\label{Future Null Infinity}

\subsection{Null infinity in the Bondi coordinate}
\label{subsecBondi}

We briefly review the essence of the asymptotic properties of the region near future null infinity
in asymptotically flat spacetimes
based on Refs.~\cite{Bondi,Sachs,Tanabe:2011es} (see also Refs. \cite{Hollands:2003ie,Hollands:2003xp,Ishibashi:2007kb}). Let $n$ be the dimension of a spacetime.
We will restrict our attention to the case $n\geq4$. 
We adopt the Bondi coordinates, 
\begin{eqnarray}
    \label{Bondi}
ds^2 = -Ae^B du^2 -2 e^B du dr + h_{IJ}r^2(dx^I + C^I du)(dx^J + C^J du),
\end{eqnarray}
where $A, B, C^I$ and $h_{IJ}$ are functions of $u$, $r$, and $x^I$. Here, $x^I$ stands for angular coordinates.
In these coordinates, future null infinity $\mathscr{I}^+$ is supposed to be located at $r=\infty$. 
Then, we expand $h_{IJ}$ near future null infinity as
\begin{eqnarray}
    \label{hass}
h_{IJ}=\omega_{IJ}+\sum_{k\geq0}h^{(k+1)}_{IJ} r^{-(n/2+k-1)},
\end{eqnarray}
where $\omega_{IJ}$ is the metric for the unit $(n-2)$-sphere,  $k\in\mathbb{Z}$ for even dimensions, and $2k\in\mathbb{Z}$ for odd dimensions.
If $h_{IJ}-\omega_{IJ}$
is nonzero, it indicates the presence of gravitational waves.
We impose the gauge condition as  
\begin{align}
    \label{gau}
    \sqrt{\det h_{IJ}}=\omega_{n-2},
  \end{align}
where $\omega_{n-2}$ is the volume element of the unit $(n-2)$-dimensional sphere.

By using the vacuum Einstein equations $R_{\mu\nu}=0$,
the falloff behavior of $A$, $B$, and $C^I$ can be given as  \cite{Tanabe:2011es}
\begin{eqnarray}
    \label{expABC}
A&=&1+ \sum^{k<n/2-2}_{k=0}A^{(k+1)}r^{-(n/2+k-1)}-m(u,x^I)r^{-(n-3)}+\Cr{-(n-5/2)},\\
B&=&B^{(1)}r^{-(n-2)}+\Cr{-(n-3/2)},\\
C^I&=& \sum^{k<n/2-1}_{k=0}C^{(k+1)I}r^{-(n/2+k)}+J^I(u,x^I)r^{-(n-1)}+\Cr{-(n-1/2)},
    \label{fallABCh}
\end{eqnarray}
where $A^{(k+1)}$, $B^{(1)}$, $C^{(k+1)I}$, $m$, and $J^I$ are functions of $u$ and $x^I$. 
    In this paper, we assume the
    above behavior of the metric
    without using the properties of field equations. 
    Hence, one may be able to apply our result to modified gravity
    theories, as well.
In general relativity, note that the integration of $m(u,x^I)$ over solid angle gives us the Bondi mass \cite{Tanabe:2011es}, 
\begin{eqnarray}
    M(u):=\frac{n-2}{16\pi}\int_{S^{n-2}}md\Omega.
   \end{eqnarray}
The nonzero components of the metric and of
the inverse metric behave as 
\begin{eqnarray}
g_{uu}&=&-Ae^B+h_{IJ}C^IC^J r^2 =-1-A^{(1)}r^{-(n/2-1)}+mr^{-(n-3)} +\Cr{-(n-1)/2}, \nonumber\\
g_{ur} &=& -e^B =-1 -B^{(1)}r^{-(n-2)} + \Cr{-(n-3/2)},\nonumber\\ 
g_{IJ} &=& h_{IJ}r^2 = \omega_{IJ}r^2 + h^{(1)}_{IJ}r^{-(n/2-3)} + \Cr{-(n-5)/2}, \nonumber \\
\quad g_{uI}&=& h_{IJ} C^J r^2 = C^{(1)}_{~~I}r^{-(n/2-2)} +\Cr{-(n-3)/2}, \nonumber \\
g^{ur} &=& -e^{-B} = -1 +B^{(1)}r^{-(n-2)} + \Cr{-(n-3/2)},\nonumber\\
g^{rr}  &=& Ae^{-B} =1+A^{(1)}r^{-(n/2-1)}-mr^{-(n-3)}+\Cr{-(n-1)/2},\nonumber\\
g^{rI} &=& C^Ie^{-B} =C^{(1)I}r^{-n/2} + \Cr{-(n+1)/2},\nonumber\\
g^{IJ} &=& h^{IJ} r^{-2} =  \omega^{IJ}\red{r^{-2}} -h^{(1)IJ}r^{-(n/2+1)} + \Cr{-(n+3)/2},
\end{eqnarray}
where
$h^{IJ}$ is defined by $h^{IJ}h_{JK}={\delta^I}_K$ and
the capital latin indices
of the quantities appearing in the right-hand side
are raised and lowered by $\omega_{IJ}$ and $\omega^{IJ}$.
In particular, in four dimensions, the behavior of
the metric components is written as
\begin{eqnarray}
    &&g_{uu}=-1+mr^{-1} +\Cr{-2}, \quad g_{ur} = -1 -B^{(1)}r^{-2} + \Cr{-3},\nonumber\\
     \quad &&g_{IJ} =  \omega_{IJ}r^2 + h^{(1)}_{IJ}r+ \Cr{0}, \quad g_{uI}= C^{(1)}_{~~I} +\Cr{-1}\nonumber \\
    &&g^{ur}= -1 +B^{(1)}r^{-2} + \Cr{-3}, \quad g^{rr} =1-mr^{-1}+\Cr{-2},\nonumber\\ \quad 
    &&g^{rI} = C^{(1)I}r^{-2} + \Cr{-3}, \quad  g^{IJ}=  \omega^{IJ}r^{-2} -h^{(1)IJ}r^{-3} + \Cr{-4},\label{gfall4d}
\end{eqnarray}
where ${\cal O}$ denotes the Landau symbol.
In Appendix~\ref{appendixa}, we present the asymptotic behavior of the Christoffel symbols.

Next, we introduce the asymptotic symmetry as the transformation which preserves the asymptotic form of the metric. Then, the variations of the components the metric \red{near} null infinity are restricted to 
\begin{eqnarray}
    \label{deltag}
    &&\delta g_{rr}=0,\quad \delta g_{rI}=0,\quad g^{IJ}\delta g_{IJ}=0, \quad\delta g_{uu}=\Cr{-(n/2-1)}, \nonumber\\
    &&\delta g_{uI}=\Cr{-(n/2-2)},\quad \delta g_{ur}=\Cr{-(n-2)},\quad \delta g_{IJ}=\Cr{-(n/2-3)},
\end{eqnarray}
where
\begin{equation}
  \delta g_{\mu\nu}:=\mbox\pounds_{\xi}g_{\mu\nu}
  \label{Def:delta-g_mu_nu}
\end{equation}
and $\xi^\mu$ is the generator of the asymptotic symmetry group \cite{Tanabe:2011es}.
Later, the asymptotic symmetry gives us
the asymptotic conserved quantities for geodesics. 

\subsection{Initial conditions of null geodesics}

In the following sections, we examine the asymptotic
behavior of null geodesics near future null infinity.
In particular, we focus on null geodesics
that correspond to worldlines of photons
emitted in the tangential directions 
to $r=\mathrm{constant}$ surfaces
near future null infinity, 
{\it i.e.},  $r'=0$, 
where the prime ($'$) denotes the derivative with respect to the affine parameter $\lambda$. 

If a black hole is present, there are null geodesics that
enter the black hole region if
its tangent vector is directed in the inward radial direction.
By contrast, \red{bearing spherical case or so in mind, one can think that} worldlines of photons emitted in the outward radial direction
 \red{would} reach future null infinity.
The null geodesics with the initial condition
given as above could have a nontrivial fate. 

%
%
%======================================%
%<<<<<<<<<<<< SECTION III  >>>>>>>>>>>>>>%
%======================================%
%
\section{asymptotic behavior of null geodesics in four dimensions}
\label{4d}

\noindent
In this section, we analyze null geodesics in four-dimensional spacetimes
and figure out sufficient conditions for spacetimes 
that any null geodesic corresponding to
the worldline of a photon emitted with 
$r^\prime = 0$ at sufficiently large $r$
reaches future null infinity.
In Sec.~\ref{subgeonull4d}, we present the geodesic equations near future null infinity
in the Bondi coordinates. Then, we analyze
the behavior of $r$ and $u$ along the null geodesics
in Secs.~\ref{4dr'}--\ref{ssestu4d}. 
The study consists of three steps.
In the first step, we show that
the geodesic has $r^{\prime\prime}>0 $
at the initial emission point (Sec.~\ref{4dr'}). Next, 
we prove that the radial coordinate $r$ of any null
geodesic with the above initial conditions will 
diverge as the affine parameter $\lambda$ is increased to infinity
(Sec.~\ref{Sec:r4d}). 
In the third step, we study the behavior of $u$
in the limit $\lambda\to \infty$ and 
prove that $u$ remains a finite value (Sec.~\ref{ssestu4d}).
This explicitly indicates that the photon
arrives at future null infinity. 
In this proof, we must require some conditions
to the property of the metric. 
These conditions are related to the presence of gravitational waves, and
indicate the possibility that photons emitted with
the above initial conditions
may not reach future null infinity without these conditions. 
In Sec.~\ref{secapcon4d}, we confirm the existence of 
the asymptotic conserved quantities for null geodesics.

\subsection{Geodesic equations and the null condition}
\label{subgeonull4d}
\noindent Here, we present the geodesic equations and the null condition in four dimensions for later convenience.
By using Eqs.~\eqref{gfall4d} and \eqref{Chfall4d}, we write down the geodesic equations near future null infinity as
\begin{eqnarray}
r'' &=&-\Gamma^{r}_{uu}{u^\prime}^2 - 2\Gamma^{r}_{ur}u'r' - \Gamma^{r}_{rr} {r^\prime}^2 - 2\Gamma^{r}_{uI} u' \left(x^I\right)^\prime  - 2\Gamma^{r}_{rI} r' \left(x^I\right)^\prime  - \Gamma^{r}_{IJ} \left(x^I\right)^\prime \left(x^J\right)^\prime  \nonumber\\
 &=& \left[ \frac12 \dot{m}r^{-1} +  \Cr{-2} \right] {u^\prime}^2
  -\Big[mr^{-2}+\Cr{-3}\Big] u'r'+\left[2B^{(1)}r^{-3}+\Cr{-4}\right] {r^\prime}^2
\nonumber\\
&& \hspace{12mm}
+\left[\left(m_{,I}-C^{(1)J}\dot{h}^{(1)}_{IJ}\right)r^{-1}+\Cr{-2}\right] u' \left(x^I\right)^\prime  -2\left[C^{(1)}_{~~I}r^{-1}+\Cr{-2}\right] r' \left(x^I\right)^\prime  \nonumber\\
&&\hspace{12mm}+
\left[ \left(\omega_{IJ}  - \frac12 \dot{h}^{(1)}_{IJ}\right) r  + \Cr{0} \right] \left(x^I\right)^\prime \left(x^J\right)^\prime   , \label{eqr4d}
\\
{u}'' &=& -\Gamma^{u}_{uu} {u^\prime}^2  -2\Gamma^{u}_{uI} u' \left(x^I\right)^\prime   -\Gamma^{u}_{IJ} \left(x^I\right)^\prime  \left(x^J\right)^\prime \nonumber\\
&=&
   \left[\left(-\dot{B}^{(1)}+\frac{1}{2}m\right)r^{-2}+\Cr{-3}\right] {u^\prime}^2  +\Cr{-2} u' \left(x^I\right)^\prime   \nonumber\\
   &&\hspace{12mm}-\left[ \omega_{IJ} r +\frac{1}{2}h^{(1)}_{IJ} +\Cr{-1}\right] \left(x^I\right)^\prime  \left(x^J\right)^\prime  ,\label{equ4d}
\end{eqnarray}
where the dot denotes the derivative with respect to $u$,
and we skipped the angular components of null geodesic equations because
we will not use them.

The null condition for the tangent vector of null geodesics,
\begin{equation}
-Ae^B {u^\prime}^2 -2 e^B u' r' + h_{IJ}r^2\left[\left(x^I\right)^\prime  + C^I u'\right]\left[\left(x^J\right)^\prime  + C^J u'\right] = 0,
\label{null4d}
\end{equation}
gives us
\begin{eqnarray}
    \label{eqnull4d}
    {u^\prime}^2&=&-2\Big[1+mr^{-1}+\Cr{-2}\Big]u'r'+\left[\omega_{IJ}r^2+\left(h^{(1)}_{IJ}+m\omega_{IJ}\right)r+\Cr{0}\right]\left(x^I\right)^\prime \left(x^J\right)^\prime \nonumber\\
    &&\hspace{12mm}+\left[2C^{(1)}_{~~I}+\Cr{-1}\right]\left(x^I\right)^\prime u'.
\end{eqnarray}
This equation can be regarded as an equation for $u^\prime$,
and the solution with a double sign
is obtained. Among them, we adopt the positive solution of $u'$ 
because we consider a future directed null geodesic.
Then, Eq.~\eqref{null4d} is \red{algebraically} solved as
\begin{multline}
  u' =
\\
  \frac{-e^B  r'+ h_{IJ} C^J r^2 \left(x^I\right)^\prime   + \sqrt{ \left[e^B r' - h_{IJ} C^J r^2 \left(x^I\right)^\prime  \right]^2 +\left(Ae^B - h_{IJ}C^I C^J r^2 \right) h_{KL}r^2 \left(x^K\right)^\prime\left(x^L\right)^\prime  } } {Ae^B - h_{MN}C^M C^Nr^2 }.
\label{u4d}
\end{multline}

\subsection{Behavior around the emission point}
\label{4dr'}

We study the behavior of $r$ of a geodesic
in the neighborhood of the emission point. 
Since $r'$ vanishes at the initial affine parameter,
$\lambda=0$, the behavior is characterized by $r''$.

For later convenience, we introduce $\left|\left(x^I\right)^\prime \right|$ as 
\begin{eqnarray}
\left|\left(x^I\right)^\prime \right| := \sqrt{\omega_{IJ} \left(x^I\right)^\prime  \left(x^J\right)^\prime }.
\end{eqnarray}
For $r'=0$, Eq.~\eqref{u4d} becomes
\begin{eqnarray}
u' &=& \frac{ h_{IJ} C^J r^2 \left(x^I\right)^\prime   + \sqrt{ \left[ h_{IJ} C^J r^2 \left(x^I\right)^\prime  \right]^2 +\left(Ae^B - h_{IJ}C^I C^J r^2 \right) h_{KL}r^2 \left(x^K\right)^\prime\left(x^L\right)^\prime  } } {Ae^B - h_{MN}C^M C^Nr^2 } \nonumber\\
&=&\Big[r+\Cr{0}\Big]  \left|\left(x^I\right)^\prime \right|.
\label{estu4d}
\end{eqnarray}
Here, note that initially $\left|\left(x^I\right)^\prime \right|\neq0$, because otherwise Eq.~\eqref{estu4d} implies  $u'=0$, that is, the tangent vector becomes zero.

At $\lambda=0$ (that is $r'=0$), Eq.~\eqref{eqr4d} becomes 
\begin{eqnarray}
    \label{r21}
r'' &=& \left[ \frac12 \dot{m}r^{-1} +  \Cr{-2} \right] {u^\prime}^2
+\left[\left(m_{,I}-C^{(1)J}\dot{h}^{(1)}_{IJ}\right)r^{-1}+\Cr{-2}\right] u' \left(x^I\right)^\prime  \nonumber\\
&&\hspace{12mm}+
\left[ \left(\omega_{IJ}   - \frac12 \dot{h}^{(1)}_{IJ}\right) r  + \Cr{0} \right] \left(x^I\right)^\prime \left(x^J\right)^\prime  \nonumber\\
&=&\Cr{-1} u' \left(x^I\right)^\prime  +  \red{\Omega_{IJ}} r \left(x^I\right)^\prime  \left(x^J\right)^\prime  + \Cr{0} \left|\left(x^I\right)^\prime \right|^2 ,
\end{eqnarray}
where we used Eq.~\eqref{eqnull4d} in the second equality, and $\Omega_{IJ}$ is defined by
\begin{eqnarray}
\Omega_{IJ} :=  \omega_{IJ}  - \frac12 \dot{h}^{(1)}_{IJ} + \frac12 \dot{m} \omega_{IJ}.
\label{assump1}
\end{eqnarray}
Furthermore with Eq.~\eqref{estu4d}, we see that the first term of the second line in the right-hand side of Eq. (\ref{r21}) is next-to-leading order and then 
\begin{eqnarray}
    \label{r2>0}
r'' = \red{\Omega_{IJ}} r \left(x^I\right)^\prime  \left(x^J\right)^\prime + \Cr{0} \left|\left(x^I\right)^\prime \right|^2.
\end{eqnarray}
The second and third terms in the expression
of Eq.~\eqref{assump1} originate from the
presence of gravitational waves. Thus, at leading order,
gravitational waves affect the null geodesic motion near
future null infinity in four dimensions.  As clarified later,
this is a fairly unique feature compared to higher-dimensional cases,
and would be related to the so-called supertranslation.
Since $\Omega_{IJ}$ does not have two positive eigenvalues
in general, one cannot claim that $r''$ is positive. 
Since ${h}^{(1)}_{IJ}$ and  $m$ appear with the $1/r$ factor in the metric, 
weak gravitational waves at large $r$ can make eigenvalues of $\Omega_{IJ}$ negative. 
The negativity of $r''$ results in the decrease in $r$ just after $\lambda=0$.
Then, there remains the possibility to have photons emitted
in the angular direction near future null infinity
which do not reach future null infinity.
Note that, in order for the value of $r$ continues to be decreased,
the sign of the eigenvalues $\Omega_{IJ}$ must be kept negative
until the velocity of the photon becomes directed toward the
central region. 
That is, gravitational waves must
be sufficiently strong or 
continuously give such an effect.  
Although such a case might be rare,
the formation of caustics of gravitational waves
at the emission point of the photon could realize such a situation.
Therefore, this result at least
indicates the importance of the effects by gravitational waves
in the dynamics of photons  
near future null infinity.
To guarantee that photons will arrive future null infinity, 
a straightforward way is to assume $\Omega_{IJ}$ to be positive definite,
which means the effects by gravitational waves are not large. 
Under this assumption, we have
\begin{eqnarray}
    \label{r2>02nd}
r'' = \red{\Omega_{IJ}} r \left(x^I\right)^\prime  \left(x^J\right)^\prime + \Cr{0} \left|\left(x^I\right)^\prime \right|^2 >0,
\end{eqnarray}
where we used $\left|\left(x^I\right)^\prime \right|\neq0$ at the initial point,
which, in turn, indicates that
there exists a constant $\lambda_c$ such that $r'>0$ for $0<\lambda<\lambda_c$. 
Then, we can prove that $r'>0$ for any positive $\lambda$
using proof by contradiction.
Suppose that there exists a positive affine parameter at which $r'=0$. Let $\lambda_0~ (>0)$ denote the minimum of such affine parameters.
Then, $r''\le0$ at this point, which gives a contradiction to Eq.~\eqref{r2>02nd}. 
Therefore, we obtain $r'> 0$ for arbitrary $\lambda>0$.  

\subsection{Asymptotic behavior of $r(\lambda)$}
\label{Sec:r4d}

In this subsection, 
we prove the existence of the lower bound of $r'$.
This implies that  
$r$ diverges as $\lambda\to \infty$.  
This result is also used for the proof of finiteness
of $u$ in the next subsection.

First, using Eq.~\eqref{u4d}, we 
estimate the order of $u'$ in terms of $\left|\left(x^I\right)^\prime \right|$. 
In the case $e^B r' - h_{IJ} C^J r^2 \left(x^I\right)^\prime   > 0$,
on the one hand, Eq.~\eqref{u4d} gives us
\begin{eqnarray}
u' &=& 
\frac{ e^B r' - h_{IJ} C^J r^2 \left(x^I\right)^\prime   }{Ae^B - h_{MN}C^M C^Nr^2}
\left[
-1+ \sqrt{1+ \frac{\left(Ae^B - h_{IJ}C^I C^J r^2 \right) h_{KL}r^2 \left(x^K\right)^\prime\left(x^L\right)^\prime  }{\left[e^B r' -  h_{PQ} C^Pr^2\left(x^Q\right)^\prime  \right]^2}}~
\right]
\nonumber \\
&\le&
 \sqrt{ \frac{  h_{IJ}r^2 \left(x^I\right)^\prime \left(x^J\right)^\prime   }{Ae^B - h_{KL}C^K C^L}r^2 }
\ =\ 
\Big[r+\Cr{0}\Big] \left|\left(x^I\right)^\prime \right|,
\label{estu4}
\end{eqnarray}
where we used $\sqrt{1+x} -1 \le \sqrt{x}$ for $x\ge 0$ in the second inequality.
On the other hand, 
in the case $e^B r' - h_{IJ} C^J r^2 \left(x^I\right)^\prime   \le 0$, we have 
\begin{eqnarray}
 \left|e^B r' - h_{IJ} C^J r^2 \left(x^I\right)^\prime \right|
  \ \le\ h_{IJ} C^J r^2 \left(x^I\right)^\prime
  \ =\ \Cr{0}\left|\left(x^I\right)^\prime\right|,
  \label{estu5_2}
\end{eqnarray}
and hence, 
Eq.~\eqref{u4d} tells us
\begin{eqnarray}
u' &=& 
\Cr{0} \left|\left(x^I\right)^\prime \right| + \sqrt{\Big[r^2+\Cr{1}\Big] \left|\left(x^I\right)^\prime \right|^2} = \Big[r+\Cr{0}\Big] \left|\left(x^I\right)^\prime \right|.
\label{estu5}
\end{eqnarray}
Therefore, we find
\begin{eqnarray}
    \label{u'4d}
u'  = \Big[r+\Cr{0}\Big] \left|\left(x^I\right)^\prime \right|
\end{eqnarray}
for arbitrary $\lambda>0$ in both cases. 
Note that $u^\prime$ is positive because the tangent vector is future directed.

By introducing positive constants ${\tilde C_1}$ and ${\tilde C}_2$, 
we can give a lower bound for $r''$ as 
\begin{eqnarray}
r'' &=& - \Big[\dot{m}r^{-1}+ \Cr{-2} \Big] u' r'  + \Big[ \Omega_{IJ} r +\Cr{0} \Big] \left(x^I\right)^\prime  \left(x^J\right)^\prime  \nonumber\\
&&\hspace{12mm}- \left[2C^{(1)}_{~~I}r^{-1}+\Cr{-2}\right] r' \left(x^I\right)^\prime  +\left[2B^{(1)}r^{-3}+\Cr{-4}\right] {r^\prime}^2 
\nonumber \\
&&\hspace{12mm}+\left[\left(C^{(1)}_{~~I}\dot{m}+m_{,I}-C^{(1)J}\dot{h}^{(1)}_{IJ}\right)r^{-1}+\Cr{-2}\right]\left(x^I\right)^\prime u'\nonumber\\
&=& - \dot{m}r^{-1} u' r'  + \Big[ \Omega_{IJ} r +\Cr{0} \Big] \left(x^I\right)^\prime  \left(x^J\right)^\prime  + \Cr{-1}r' \left|\left(x^I\right)^\prime \right| \nonumber\\
&&\hspace{12mm}+ \left[2B^{(1)}r^{-3}+\Cr{-4}\right] {r^\prime}^2
\nonumber \\
&>& - \dot{m}r^{-1} u' r'  + \Big[ \Omega_{IJ} r +\Cr{0} \Big] \left(x^I\right)^\prime  \left(x^J\right)^\prime   -  \tilde C_1r^{-1} r' \left|\left(x^I\right)^\prime \right| + \left[2B^{(1)}r^{-3}+\Cr{-4}\right] {r^\prime}^2 
\nonumber\\
&\geq& - \dot{m}r^{-1} u' r'  + \Big[ \Omega_{IJ} r +\Cr{0} \Big] \left(x^I\right)^\prime  \left(x^J\right)^\prime  - \frac{1}{2}\tilde C_1\left[r^{-2}{r^\prime}^2 +\left|\left(x^I\right)^\prime \right|^2\right]+\left[2B^{(1)}r^{-3}+\Cr{-4}\right] {r^\prime}^2 \nonumber\\
&>& - \dot{m}r^{-1} u' r'  + \Big[ \Omega_{IJ} r +\Cr{0} \Big] \left(x^I\right)^\prime  \left(x^J\right)^\prime  - \tilde C_2r^{-2}{r^\prime}^2,
\label{estr2}
\end{eqnarray}
where we used Eqs.~\eqref{eqr4d} and~\eqref{eqnull4d} in the first equality,
used Eq.~\eqref{u'4d} in the second equality,
gave a lower bound for the coefficient of $r'\left|\left(x^I\right)^\prime \right|$ in the third inequality,
used the arithmetic-geometric mean inequality 
\begin{eqnarray}
\left| r^{-1} r' \left|\left(x^I\right)^\prime \right| \right| \le \frac{1}{2}\left[r^{-2} {r^\prime}^2  +  \left|\left(x^I\right)^\prime \right|^2  \right]
\end{eqnarray}
in the fourth inequality, and gave a lower bound for the coefficient of ${r^\prime}^2$ in the fifth inequality.

The Einstein equation implies the monotonicity of $m(u\red{,x^I})$
as $\dot{m}=-\frac14 \dot h^{(1)}_{IJ} \dot h^{(1)IJ}\red{ \leq 0}$ \red{\cite{Tanabe:2011es}}.
Since this is a natural property for $m(u\red{,x^I})$ regardless of gravitational theories,
we assume this monotonicity for $m(u\red{,x^I})$, 
\begin{eqnarray}
   \dot{m} \leq 0.
\label{assump2}
\end{eqnarray}
Under the assumptions of the positive definiteness of $\Omega_{IJ}$  and \eqref{assump2}, we have
\begin{eqnarray}
r''&>& - \tilde C_2r^{-2} {r^\prime}^2,
\label{estr2f}
\end{eqnarray}
from Eq.~\eqref{estr2}.
Inequality~\eqref{estr2f} and the positivity of $r'$ give 
    \begin{eqnarray}
        \frac{r''}{r'} > - \frac{{\tilde C}_2}{r^2} r'.
    \end{eqnarray}
By integrating out this inequality, 
    \begin{eqnarray}
        \log r' > \frac{{\tilde C}_2}{r}+{\tilde C}_3
    \end{eqnarray}
is obtained, where ${\tilde C}_3$ is the integral constant.
Thus, we have
    \begin{eqnarray}
        r' &>& \exp\left({\frac{{\tilde C}_2}{r}+{\tilde C}_3}\right)
        \ >\ {\tilde C}_4,
    \end{eqnarray}
where ${\tilde C}_4:=e^{{\tilde C}_3} >0$.
Integrating this inequality again, we obtain
    \begin{eqnarray}
        \label{r>4d}
        r &>& {\tilde C}_4 \lambda + {\tilde C}_5,\label{estfinr4d}
    \end{eqnarray}
where ${\tilde C}_5$ is the integral constant.
Thus, $r$ diverges as $\lambda\to\infty$.
Note that the same procedure does not work without
the assumption of Eq.~\eqref{assump2}
because the term $- \dot{m}r^{-1} u' r' $ in the last line of Eq.~\eqref{estr2} gives the contribution of $\Cr{0}r'\left|\left(x^I\right)^\prime \right|$
through Eq.~\eqref{u'4d}.

To confirm that the current null geodesics reach
future null infinity, one has to check that
$u(\lambda)$ asymptotically converges to a finite value.
We study this issue in the next subsection. 

\subsection{Asymptotic behavior of $u(\lambda)$} 
\label{ssestu4d}

We now examine the asymptotic behavior of $u(\lambda)$. Equation~\eqref{eqnull4d} gives 
\begin{eqnarray}
&&\hspace{-12mm}\left[\omega_{IJ}+\left(h^{(1)}_{IJ}+m\omega_{IJ}\right)r^{-1}+\Cr{-2}\right]\left(x^I\right)^\prime \left(x^J\right)^\prime 
\nonumber \\ 
&
=&
r^{-2}{u^\prime}^2+2\Big[ r^{-2}+mr^{-3}+\Cr{-4}\Big] u'r'-\left[2C^{(1)}_{~~I}r^{-2}+\Cr{-3}\right]\left(x^I\right)^\prime u' 
\nonumber \\ 
&
=&\Big[ r^{-2} +\Cr{-3} \Big] {u^\prime}^2 
+2\Big[ r^{-2} +mr^{-3}+ \Cr{-4} \Big] u' r' + \Cr{-1}  {\left|\left(x^I\right)^\prime \right|}^2 ,
\end{eqnarray}
where the first equality is obtained in a simple rearrangement of Eq.~\eqref{eqnull4d} and 
we used the arithmetic-geometric mean inequality  
\begin{eqnarray}
0\le r^{-2} u' \left|\left(x^I\right)^\prime \right| =  \left(r^{-3}{u^\prime}^2\right)^{1/2} \left[r^{-1}\left|\left(x^I\right)^\prime \right|^2\right]^{1/2} \le 
\frac{1}{2}\left[r^{-3}{u^\prime}^2 + r^{-1}\left|\left(x^I\right)^\prime \right|^2\right],
\end{eqnarray}
which implies
\begin{eqnarray}
    \label{agiux4d}
r^{-2}u' \left(x^I\right)^\prime  = \Cr{-3} {u^\prime}^2 + \Cr{-1} \left|\left(x^I\right)^\prime \right|^2
\end{eqnarray}
in the order estimate of the second equality. 
This means 
\begin{eqnarray}
    \label{nconome4d}
\Big[ \omega_{IJ} + \Cr{-1} \Big] \left(x^I\right)^\prime  \left(x^J\right)^\prime  
= \Big[ r^{-2} + \Cr{-3} \Big] {u^\prime}^2 
+\Big[ 2r^{-2} + \Cr{-3} \Big] u' r' .
\end{eqnarray}
Substituting this relation into Eq.~\eqref{equ4d}, we have 
\begin{eqnarray}
{u}'' 
&=&
\left[\left(-\dot{B}^{(1)}+\frac{1}{2}m\right)r^{-2}+\Cr{-3}\right] {u^\prime}^2  +  \Cr{-3} {u^\prime}^2 + \Cr{-1} {{\left|\left(x^I\right)^\prime \right|}}^2 \nonumber\\
&&\hspace{12mm}-\left[\omega_{IJ} r +\frac{1}{2}h^{(1)}_{IJ} +\Cr{-1}\right] \left(x^I\right)^\prime  \left(x^J\right)^\prime  
\nonumber \\
&=&
\Cr{-2} {u^\prime}^2  - \Big[ \omega_{IJ}  + \Cr{-1} \Big]r \left(x^I\right)^\prime  \left(x^J\right)^\prime 
\nonumber \\
&=&
\Cr{-2} {u^\prime}^2
-\left[ \frac{1}{r} + \Cr{-2} \right] {u^\prime}^2 
-\left[ \frac{2}{r} + \Cr{-2} \right] u' r' 
\nonumber \\
&=&-\left[ \frac{1}{r} + \Cr{-2} \right] {u^\prime}^2 
-\left[ \frac{2}{r} + \Cr{-2} \right] u' r' 
\nonumber \\
&<&-\left( \frac{2}{r} - \frac{{\tilde C_6}}{r^2} \right) u' r' \label{u2}
\end{eqnarray}
for large $r$ ({\it i.e.} for large $\lambda$),
where we used Eq.~\eqref{agiux4d} in the first equality,
used Eq.~\eqref{nconome4d} in the third equality,
and excluded a nonpositive term and gave an upper bound for
the coefficient of $u'r'$ in the fifth inequality
by introducing a positive constant ${\tilde C_6}$.
Next, we define $U$ as 
\begin{eqnarray}
  U:= r^2 \exp\left({\frac{{\tilde C_6}}{r}}\right) u',
  \label{Def:U}
\end{eqnarray}
and then, the above inequality is simply written as
\begin{eqnarray}
U'<0.
\end{eqnarray}
Integration of this inequality gives
\begin{eqnarray}
U  <  {\tilde C_7},
\end{eqnarray}
where ${\tilde C_7}$ is a positive constant. 
Recalling the definition of $U$ of Eq.~\eqref{Def:U}, we have 
\begin{eqnarray}
0\le u'  < {\tilde C_7} r^{-2} < {\tilde C_7} \left({\tilde C}_4 \lambda + {\tilde C}_5\right)^{-2},  \label{u'est}
\end{eqnarray}
where $\exp\left({-{{\tilde C_6}}/{r}}\right)<1$ was used.
Integrating this inequality in the domain $[\lambda_{\rm L}, \lambda]$, we have 
\begin{eqnarray}
u-u|_{\lambda=\lambda_{\rm L}} < - \frac{\tilde C_7}{{\tilde C}_4}  \left[ \left({\tilde C}_4 \lambda + {\tilde C}_5\right)^{-1} -\left({\tilde C}_4 \lambda_{\rm L} + {\tilde C}_5\right)^{ -1}\right].
\end{eqnarray}
Therefore, $u$ is bounded from above as
\begin{eqnarray}
u<\frac{\tilde C_7}{{\tilde C}_4} \left({\tilde C}_4 \lambda_{\rm L} + {\tilde C}_5\right)^{ -1} + 
u|_{\lambda=\lambda_{\rm L}},
\end{eqnarray}
and thus, $u$ does not diverge. Therefore, the null geodesic reaches
future null infinity under the assumptions of the positive definiteness
of $\Omega_{IJ}$ and Eq.~\eqref{assump2}.
We stress that these assumptions are not so strong in realistic
physical processes. In addition, this conclusion holds
for any null geodesics with $r'(0)\ge 0$.

\subsection{Asymptotic constants of motion} 
\label{secapcon4d}

Since asymptotically flat spacetimes have the asymptotic symmetry
as explained at the end of Sec.~\ref{subsecBondi},
we expect that a geodesic has constants of motion
in approximate sense near future null infinity.
We confirm this here. 
Let $\xi$ be a generator of the asymptotic symmetry. 
We define $Q_\xi$ by
\begin{equation}
  Q_\xi\ :=\ ({x^\mu})'\xi_\mu.
\end{equation}
Recalling the definition of $\delta g_{\mu\nu}$
given in Eq.~\eqref{Def:delta-g_mu_nu}, 
the derivative of $Q_\xi$ with respect to the affine parameter of this geodesic is calculated as
\begin{eqnarray}
    \left(x^\mu\right)^\prime \nabla_\mu Q_\xi&=&\frac{1}{2}\left(x^\mu\right)^\prime  \left(x^\nu\right)^\prime \mbox\pounds_{\xi}g_{\mu\nu}\nonumber\\
    &=&\frac{1}{2}\left[{u^\prime}^2\delta g_{uu}+2u'r'\delta g_{ur}+\left(x^I\right)^\prime \left(x^J\right)^\prime \delta g_{IJ}+2u'\left(x^I\right)^\prime \delta g_{uI}\right]\nonumber\\
    &=&\frac{1}{2}\left[{u^\prime}^2\Cr{-1}+2u'r'\Cr{-2}+\left(x^I\right)^\prime \left(x^J\right)^\prime \Cr{1}+2u'\left(x^I\right)^\prime \Cr{0}\right]\nonumber\\
    &=&\frac{1}{2}\Big[\Cr{-5}+\Cr{-4}+\Cr{-3}+\Cr{-4}\Big]\nonumber\\
    &=&\Cr{-3}, \label{apcon4d}
\end{eqnarray}
where we used Eq.~\eqref{deltag} in the third equality and 
used Eqs.~\eqref{u'est},~\eqref{r'<}, and~\eqref{xI<} in the fourth equality.
This can be regarded as the approximate conservation law
because 
\begin{eqnarray}
    Q_\xi&=&{\rm constant}+\Cr{-2}
\end{eqnarray}
holds from Eqs.~\eqref{r>4d} and~\eqref{r<}.
In the case $\xi=\partial_u$, Eq.~\eqref{apcon4d} corresponds to
the approximate conservation of the energy.
There also exists the Killing vector $\xi=f^I\partial_I$
that represents rotational symmetry, and for this choice,
Eq.~\eqref{apcon4d} corresponds to the approximate conservation
of the angular momentum.

%
%
%======================================%
%<<<<<<<<<<<< SECTION IV  >>>>>>>>>>>>>>%
%======================================%
%
\section{asymptotic behavior of null geodesics in Higher dimensions}
\label{highd}

In this section, we study null geodesics in higher dimensions $n\geq5$
paying attention to the difference from the case $n=4$.
Although most of the analyses in this section are
parallel to that in Sec.~\ref{4d},
a critical difference arises in the power of $r$.
In particular, it is shown that 
any null geodesic with the initial condition $r'=0$ at sufficiently large $r$
reaches future null infinity without any additional assumption. 
First, we show the geodesic equations near future null infinity
in Sec.~\ref{subgeonullhd}. Next, we discuss the behavior of $r$
along the null geodesics emitted with $r'=0$
in Secs.~\ref{hdr'} and \ref{Sec:rhd}. Then,
we prove the finiteness of $u$ along the null geodesics
in Sec.~\ref{ssestuhd}. Last, we confirm the
existence of asymptotic conserved quantities for null geodesics
in Sec.~\ref{secapconhd}. 

\subsection{Geodesic equations and the null condition}
\label{subgeonullhd}

In this subsection, we present the geodesic equations and the null condition for $n\geq5$. The geodesic equations near null infinity are
\begin{eqnarray}
r'' &=&  \left[-\frac{1}{2}\dot{A}^{(1)}r^{-(n/2-1)}+ \Cr{- (n-1)/2}\right] {u^\prime}^2+\left[\frac{n-2}{2}A^{(1)}r^{-n/2}+\Cr{-(n+1)/2}\right] u'r'\nonumber\\
&&\hspace{12mm}+\left[(n-2)B^{(1)}r^{-(n-1)}+\Cr{-(n-1/2)}\right] {r^\prime}^2\nonumber\\
&&\hspace{12mm}+\left[\left(-\frac{n-4}{2}C^{(1)}_{~~I}-A^{(1)}_{,I}\right)r^{-(n/2-1)}+\Cr{-(n-1)/2}\right] u' \left(x^I\right)^\prime  \nonumber\\
&&\hspace{12mm}
-\left[\frac{n}{2}C^{(1)}_{~~I}r^{-(n/2-1)}+\Cr{-(n-1)/2}\right] r' \left(x^I\right)^\prime  \nonumber\\
&&\hspace{12mm}+
\left[\omega_{IJ} r-\frac{1}{2}\dot{h}^{(1)}_{IJ}r^{-(n/2-3)} +\Cr{-(n-5)/2} \right] \left(x^I\right)^\prime \left(x^J\right)^\prime   , \label{eqrhd}
\\
{u}'' &=&   -\left[\frac{n-2}{4}A^{(1)}r^{-n/2}+\Cr{-(n+1)/2)}\right] {u^\prime}^2  \nonumber\\
&&\hspace{12mm}-2\left[-\frac{n-4}{4}C^{(1)}_{~~I}r^{-(n/2-1)}+ \Cr{-(n-1)/2}\right] u' \left(x^I\right)^\prime   \nonumber\\
&&\hspace{12mm}- \left[ \omega_{IJ} r -\frac{n-6}{4}h^{(1)}_{IJ}r^{-(n/2-2)} +\Cr{-(n-3)/2}\right] \left(x^I\right)^\prime  \left(x^J\right)^\prime  ,\label{equhd}
\end{eqnarray}
where we skipped the angular component of null geodesic equations
because we will not use them.
The null condition for the tangent vector of null geodesics
is the same as Eq.~\eqref{null4d}.
For general $n\ge 5$, Eq.~\eqref{null4d} gives 
\begin{eqnarray}
    \label{eqnullhd}
    {u^\prime}^2&=&-2\left[1-A^{(1)}r^{-(n/2-1)}+\Cr{-(n-1)/2}\right]u'r'\nonumber\\
    &&\hspace{12mm}+\left[\omega_{IJ}r^2+\left(h^{(1)}_{IJ}-\omega_{IJ}A^{(1)}\right)r^{-(n/2-3)}+\Cr{-(n-5)/2}\right]\left(x^I\right)^\prime \left(x^J\right)^\prime \nonumber\\
    &&\hspace{12mm}+\left[2C^{(1)}_{~~I}r^{-(n/2-2)}+\Cr{-(n-3)/2}\right]\left(x^I\right)^\prime u'.
\end{eqnarray}
The \red{algebraic} solution for $u^\prime$ is the same as Eq.~\eqref{u4d}.

\subsection{Behavior around the emission point}
\label{hdr'}

In this subsection, we show $r'>0$ after the emission
by the parallel argument to Sec.~\ref{4dr'}.
We focus on the case where $r'$ is initially zero and
set $\lambda=0$ at the emission point. 
Then, the relation
$u^\prime = \Cr{1}  \left|\left(x^I\right)^\prime \right| $,
given in Eq.~\eqref{estu4d}, 
holds also for $n\ge 5$.
Equation~\eqref{eqrhd} at $\lambda=0$ becomes
\begin{eqnarray}
    \label{etr2hd}
r'' = \Cr{-(n/2-1)} u' \left(x^I\right)^\prime  + \omega_{IJ} r \left(x^I\right)^\prime  \left(x^J\right)^\prime  + \Cr{-(n/2-3)} \left|\left(x^I\right)^\prime \right|^2 ,
\end{eqnarray}
where $\left|\left(x^I\right)^\prime \right|\neq0$ as discussed in Sec.~\ref{4dr'}.
Furthermore, with Eq.~\eqref{estu4d},
the first term in the right-hand side is of higher order, and then, 
\begin{eqnarray}
r'' = \omega_{IJ} r \left(x^I\right)^\prime  \left(x^J\right)^\prime + \Cr{-(n/2-3)} \left|\left(x^I\right)^\prime \right|^2 >0
\end{eqnarray}
holds. 
This equation is different from Eq.~\eqref{r2>0} in
the four-dimensional case:
neither $\dot{h}^{(1)}_{IJ}$ nor $\dot{m}$ is included in the coefficient of $r$.
This is because the falloff of the metric is faster in higher dimensions,
which is the same reason why the supertranslation group
and the memory effect are absent in higher dimensions \cite{Tanabe:2011es, Hollands:2016oma}. 
By the same argument as that of Sec.~\ref{4dr'} after Eq.~\eqref{r2>02nd},
we obtain $r'>0$ for arbitrary $\lambda>0$. 

\subsection{Asymptotic behavior of $r(\lambda)$}
  \label{Sec:rhd}

We now consider the asymptotic behavior of $r(\lambda)$ for $\lambda\to \infty$
along the null geodesics.
In a similar manner to the study in Sec.~\ref{Sec:r4d}, 
we relate  $u'$ to $\left|\left(x^I\right)^\prime \right|$. 
In the case $e^B r' - h_{IJ} C^J r^2 \left(x^I\right)^\prime   > 0$, on the one hand, 
the calculation similar to Eq.~\eqref{estu4} gives $u'  = \Big[r+\Cr{-(n/2-2)}\Big] \left|\left(x^I\right)^\prime \right|$.
In the case $e^B r' - h_{IJ} C^J r^2 \left(x^I\right)^\prime   \le 0$, on the other hand, we obtain the same equation as Eq.~\eqref{estu5_2}
but with $\Cr{0}$ of the right-hand side 
being replaced by $\Cr{-(n/2-2)}$, and 
Eq.~\eqref{u4d} gives us
\begin{eqnarray}
u' &=& 
\Cr{-(n/2-2)} \left|\left(x^I\right)^\prime \right| + \sqrt{\Big[r^2+\Cr{-(n/2-3)}\Big] \left|\left(x^I\right)^\prime \right|^2} \nonumber\\
&=& \left[r+\Cr{-(n/2-2)}\right] \left|\left(x^I\right)^\prime \right|.
\label{estu3}
\end{eqnarray}
Therefore, we have $u'  = \Big[r+\Cr{-(n/2-2)}\Big] \left|\left(x^I\right)^\prime \right|$,
the leading order being the same as that of Eq.~\eqref{u'4d}, 
for arbitrary $\lambda>0$ in both cases. 

We now consider Eq.~\eqref{eqrhd}. 
$r''$ is calculated as
\begin{eqnarray}
r''  &=&\left[\dot{A}^{(1)}r^{-(n/2-1)}+\Cr{-(n-1)/2}\right]u'r'\nonumber\\
&&\hspace{12mm}+\left[\omega_{IJ}r-\frac{1}{2}\left(\dot{A}^{(1)}\omega_{IJ}+\dot{h}^{(1)}_{IJ}\right)r^{-(n/2-3)}+\Cr{-(n-5)/2} \right]\left(x^I\right)^\prime  \left(x^J\right)^\prime \nonumber\\
&&\hspace{12mm}+\left[-\left(\frac{n-4}{2}C^{(1)}_{~~I}+A^{(1)}_{,I}\right)r^{-(n/2-1)}+\Cr{-(n-1)/2}\right]u'\left(x^I\right)^\prime  \nonumber\\
&&\hspace{12mm}+\left[(n-2)B^{(1)}r^{-(n-1)}+\Cr{-(n-1/2)}\right] {r^\prime}^2\nonumber\\
&&\hspace{12mm}-\left[\frac{n}{2}C^{(1)}_{~~I}r^{-(n/2-1)}+\Cr{-(n-1)/2}\right]r'\left(x^I\right)^\prime \nonumber\\
&=& \left[\omega_{IJ}r-\frac{1}{2}\left(\dot{A}^{(1)}\omega_{IJ}+\dot{h}^{(1)}_{IJ}\right)r^{-(n/2-3)}+\Cr{-(n-5)/2} \right]\left(x^I\right)^\prime  \left(x^J\right)^\prime \nonumber\\
&&\hspace{12mm}  +\Cr{-(n/2-2)} r' \left|\left(x^I\right)^\prime \right|
+\left[(n-2)B^{(1)}r^{-(n-1)}+\Cr{-(n-1/2)}\right] {r^\prime}^2,
\end{eqnarray}
where we used Eqs.~\eqref{eqrhd} and~\eqref{eqnullhd} in the first equality 
and used Eq.~\eqref{u'4d} in the second equality.
In contrast to the four-dimensional case,
the condition of Eq.~\eqref{assump2}
is not necessary because the corresponding term $-\dot{m}r^{-(n-3)}u'r'$ is of
higher order.
For a technical reason, 
we introduce $\alpha$ that satisfies $0<\alpha<1$.
Then, we have
\begin{eqnarray}
r''&>& \left( \omega_{IJ}r  +\Cr{\alpha} \right) \left(x^I\right)^\prime  \left(x^J\right)^\prime  - {\hat C}_1r^{-(n+\alpha-4)}{r^\prime}^2
\nonumber \\
&\geq& - {\hat C}_1r^{-(n+\alpha-4)} {r^\prime}^2,
\label{estr3}
\end{eqnarray}
where we used the arithmetic-geometric mean inequality 
\begin{eqnarray}
\left| r^{-(n/2-2 )} r' \left|\left(x^I\right)^\prime \right| \right| \le \frac{1}{2}\left[r^{-(n+\alpha-4)} {r^\prime}^2  +  r^{\alpha}\left|\left(x^I\right)^\prime \right|^2\right]
\end{eqnarray}
in the first line, we used $\alpha<1$ in the second line, and ${\hat C}_1$ is a positive constant.
Here, the introduction of $\alpha$ is necessary for the case
$n=5$, and we can set $\alpha=0$ for $n\ge 6$. 
Equation~\eqref{estr3} and the positivity of $r'$
implies $r'^{-1}r'' > - {\hat C}_1r^{-(n+\alpha-4)} r'$, 
and integrating this inequality, we obtain
    \begin{eqnarray}
        \log r' > \frac{{\hat C}_1}{n+\alpha-5}r^{-(n+\alpha-5)}+{\hat C}_2,
    \end{eqnarray}
where ${\hat C}_2$ is an integral constant.
Then, we have
    \begin{eqnarray}
        r' &>& \exp\left(\frac{{\hat C}_1}{n+\alpha-5}r^{-(n+\alpha-5)}+{\hat C}_2\right)\ >\ e^{{\hat C}_2},
    \end{eqnarray}
    where we used $\alpha>0$ and $n\geq5$.
    Integrating this inequality again, we obtain
    \begin{eqnarray}
        \label{r>hd}
        r &>& e^{{\hat C}_2} \lambda + {\hat C}_3,\label{estfinrhd}
    \end{eqnarray}
    where ${\hat C}_3$ is an integral constant.
    Thus, $r$ diverges to infinity as $\lambda\to\infty$.

\subsection{Asymptotic behavior of $u(\lambda)$} 
\label{ssestuhd}

Here, we examine the asymptotic behavior of $u(\lambda)$ as in
a similar manner to Sec.~\ref{ssestuhd}. Equation~\eqref{eqnullhd} gives 
    \begin{eqnarray}
        &&\hspace{-10mm}\left[\omega_{IJ}+\left(h^{(1)}_{IJ}-\omega_{IJ}A^{(1)}\right)r^{-(n/2-1)}+\Cr{-(n-1)/2}\right]\left(x^I\right)^\prime \left(x^J\right)^\prime \nonumber\\
        &=&r^{-2} {u^\prime}^2 +2\left[r^{-2}-A^{(1)}r^{-(n/2+1)}+\Cr{-(n+3)/2}\right]u'r'\nonumber\\
        &&\hspace{12mm}-\left[2C^{(1)}_{~~I}r^{-n/2}+\Cr{-(n+1)/2}\right]\left(x^I\right)^\prime u'\nonumber\\
        &=&r^{-2} {u^\prime}^2 +2\left[r^{-2}-A^{(1)}r^{-(n/2+1)}+\Cr{-(n+3)/2}\right]u'r'\nonumber \\ &&  \hspace{12mm}+\Cr{-3} {u^\prime}^2 + \Cr{-(n-3)}  {\left|\left(x^I\right)^\prime \right|}^2 ,
        \end{eqnarray}
where we used the arithmetic-geometric mean inequality 
\begin{eqnarray}
0\le r^{-n/2} u' \left|\left(x^I\right)^\prime \right| =  \left(r^{-3}{u^\prime}^2\right)^{1/2} \left[{r^{-(n-3)}}{\left|\left(x^I\right)^\prime \right|}^2\right]^{1/2} \le 
\frac{1}{2}\left[r^{-3}{u^\prime}^2 + r^{-(n-3)}{\left|\left(x^I\right)^\prime \right|}^2\right],
\end{eqnarray}
that implies
\begin{eqnarray}
    \label{agiuxhd}
r^{-n/2} u' \left(x^I\right)^\prime  = \Cr{-3} {u^\prime}^2 + \Cr{-(n-3)}  {\left|\left(x^I\right)^\prime \right|}^2
\end{eqnarray}
to estimate the order of terms in the second equality.
This indicates 
\begin{eqnarray}
    \label{nconomehd}
    %&\hspace{-30mm}
    \left[\omega_{IJ}
    %+\left(h^{(1)}_{IJ}-\omega_{IJ}A^{(1)}\right)r^{-(n/2-1)}+\Cr{-(n-1)/2}
    +\Cr{-(n/2-1)}\right]\left(x^I\right)^\prime \left(x^J\right)^\prime
    %\nonumber\\
%&
= \Big[r^{-2} + \Cr{-3} \Big] {u^\prime}^2 
+2\Big[r^{-2}%-A^{(1)}r^{-(n/2+1)}+\Cr{-(n+3)/2}
+\Cr{-(n/2+1)}\Big]u'r' .
\end{eqnarray}
Substituting this into Eq.~\eqref{equhd}, we obtain 
\begin{eqnarray}
    {u}'' &=&   -\left[\frac{n-2}{4}A^{(1)}r^{-n/2}+\Cr{-(n+1)/2)}\right] {u^\prime}^2 +  \Cr{-2} {u^\prime}^2 + \Cr{-(n-4)} {{\left|\left(x^I\right)^\prime \right|}}^2 \nonumber\\
    &&\hspace{12mm}- \left[ \omega_{IJ} r -\frac{n-6}{4}h^{(1)}_{IJ}r^{-(n/2-2)} +\Cr{-(n-3)/2}\right] \left(x^I\right)^\prime  \left(x^J\right)^\prime \nonumber\\
    &=&  \Cr{-2} {u^\prime}^2 - \left[ \omega_{IJ}+\Cr{-(n/2-1)}  \right]r \left(x^I\right)^\prime  \left(x^J\right)^\prime \nonumber\\
&=&
\Cr{-2} {u^\prime}^2
-\Big[ r^{-1} + \Cr{-2} \Big] {u^\prime}^2 
-\left[ 2r^{-1} + \Cr{-n/2} \right] u' r' 
\nonumber \\
&=&-\Big[ r^{-1} + \Cr{-2} \Big] {u^\prime}^2 
-\left[ 2r^{-1} + \Cr{-n/2} \right] u' r' 
\nonumber \\
&<&-\left( \frac{2}{r} - \frac{{\hat C_4}}{r^2} \right) u' r' \label{u2hd}
\end{eqnarray}
for large $r$ ({\it i.e.} for large $\lambda$),
where we used Eq.~\eqref{agiuxhd} in the first equality,
used Eq.~\eqref{nconomehd} in the third equality,
and excluded a nonpositive term and gave an upper bound for the coefficient of $u'r'$ in the fifth inequality by introducing a positive constant ${\hat C_4}$.
Thus, by the same argument as in Sec.~\ref{ssestu4d} after Eq.~\eqref{u2}, 
\begin{eqnarray}
    \label{u'esthd}
u'=\Cr{-2}=\cal{O}(\lambda \rm{^{-2})},
\end{eqnarray}
and $u$ does not diverge. Therefore,
the null geodesic reaches future null infinity.
Again, this conclusion is also correct for any null geodesic with $r'(0)>0$. 

\subsection{Asymptotic constants of motion} 
\label{secapconhd}

It is expected that a geodesic has constants of motion
in the approximate sense near future null infinity
in higher-dimensional spacetimes as well. 
In this subsection, we show the approximately conserved quantities using the results in Appendix~\ref{appendixb}.

In a similar manner to Eq.~\eqref{apcon4d},
the derivative of $Q_\xi$ with respect to $\lambda$ is
\begin{eqnarray}
    \left(x^\mu\right)^\prime \nabla_\mu Q_\xi
    &=&\frac{1}{2}\left[{u^\prime}^2\delta g_{uu}+2u'r'\delta g_{ur}+\left(x^I\right)^\prime \left(x^J\right)^\prime \delta g_{IJ}+2u'\left(x^I\right)^\prime \delta g_{uI}\right]\nonumber\\
    &=&\frac{1}{2}\left[{u^\prime}^2\Cr{-(n/2-1)}+2u'r'\Cr{-(n-2)}+\left(x^I\right)^\prime \left(x^J\right)^\prime \Cr{-(n/2-3)}\nonumber\right.\\
    &&\left.\hspace{12mm}+2u'\left(x^I\right)^\prime \Cr{-(n/2-2)}\right]\nonumber\\
    &=&\frac{1}{2}\left[\Cr{-(n/2+3)}+\Cr{-n}+\Cr{-(n/2+1)}+\Cr{-(n/2+2)}\right]\nonumber\\
    &=&\Cr{-(n/2+1)},\label{apconhd}
\end{eqnarray}
where we used Eq.~\eqref{deltag} in the second equality and Eqs.~\eqref{u'est},~\eqref{r'<}, and \eqref{xI<} in the third equality.
This can be regarded as the approximate conservation law. From Eqs.~\eqref{r>hd} and~\eqref{r<}, we have 
\begin{eqnarray}
    Q_\xi&=&{\rm constant}+\Cr{-n/2}.
\end{eqnarray}
In the case $\xi=\partial_u$, Eq.~\eqref{apconhd} corresponds to the approximate conservation of the energy. In the case that $\xi$ can be written as $\xi=f^I\partial_I$, Eq.~\eqref{apconhd} corresponds to the approximate conservation of the angular momentum.

%
%
%======================================%
%<<<<<<<<<<<< SECTION V  >>>>>>>>>>>>>>%
%======================================%
%
\section{Conclusions}
\label{summary}

In this paper, we have analyzed null geodesics
that correspond to worldlines of photons emitted with the initial condition
$r'=0$ (or $r'>0$) at which $r$ is sufficiently large
in the Bondi coordinates.
We have proven that any such geodesic 
reaches future null infinity under the asymptotically
flat conditions in the higher-dimensional cases.
In the four-dimensional cases, the additional assumptions have been required to be imposed. 
There is a nontrivial difference between the cases
in four dimensions and in higher dimensions.

The  two assumptions required in the four-dimensional cases are 
the positive definiteness of $\Omega_{IJ}\red{:=}\omega_{IJ}  - \frac12 \dot{h}^{(1)}_{IJ} + \frac12 \dot{m} \omega_{IJ}$ and $\dot{m}\leq0$.
The latter condition is satisfied in general relativity. 
The former condition is not trivial,
and it is satisfied if the effects of gravitational waves
are sufficiently weak near future null infinity. 
Although the case where the null geodesic does not
reach future null infinity might be  rare, 
there is a possibility that the null geodesic may not reach
future null infinity if we tune the gravitational wave emission
(e.g., formation of caustics just at the emission point). 
In the case of higher dimensions, by contrast, 
these assumptions are not necessary for any null geodesic
to reach future null infinity. \red{In future work, it should be clarified whether the sufficient condition in four dimensions is also the necessary condition or not.}

It should be noted that under the assumption that $\dot{m}\leq0$
in four dimensions 
the positive definiteness of $\Omega_{IJ}$ is a stronger condition
than the positive definiteness of $\chi_{IJ} = -\Gamma^r_{~IJ} = \omega_{IJ} r  -  \frac12 \dot{h}^{(1)}_{IJ} r + \Cr{0}$,
where $\chi_{IJ}$ is the extrinsic curvature of $r=\mathrm{constant}$ hypersurface
in the $u=\mathrm{constant}$ subspace. 
It should be beneficial to investigate the meaning of $\Omega_{IJ}$
in more detail. The difference between the four-dimensional cases and
higher-dimensional cases motivates us to 
investigate the relation 
to the memory effect, which also provides a nontrivial difference
between four dimensions and higher dimensions due to the asymptotic
behavior of the metric. The interpretation of $\Omega_{IJ}$ would serve
as a key to understand the connection of our analysis with the memory effect.

The study of this paper is the first step toward clarifying
the general properties of
the global behavior of photons in general dynamical spacetimes.
One of the possible applications is to characterize the
strong gravity regions by extending the concepts of photon sphere
from a global point of view, while most of the existing
generalizations of the photon sphere 
are defined locally or in spacetimes with symmetries,
or have difficulty in specifying it by calculation~\cite{Shiromizu:2017ego,Yoshino:2017gqv,Siino:2019vxh,Yoshino:2019dty,Cao:2019vlu}.
It is also interesting to relate such study to the observation
of the black hole shadow because it might become possible to
observe the neighborhood of dynamically evolving black holes
in the near future.

%
%======================================%
%<<<<<<<<<<<< acknowledgments  >>>>>>>>>>>>>>%
%======================================%
%

\acknowledgments

M. A. is grateful to Professor S. Mukohyama and Professor T. Tanaka for continuous encouragements and useful suggestions. K. I. and T. S. are supported by Grant-Aid for Scientific Research from Ministry of Education, Science, Sports and Culture of Japan (Grant No. JP17H01091). 
K.~I. is also supported by JSPS Grants-in-Aid for Scientific Research (B) (Grant No. JP20H01902).
T. S. is also supported by JSPS Grants-in-Aid for Scientific Research (C) (Grant No. JP21K03551). H.~Y. is in part supported by JSPS KAKENHI Grants No.
JP17H02894 and JP18K03654. The work of H.Y. is partly supported 
by Osaka City University Advanced Mathematical Institute 
(MEXT Joint Usage/Research Center on Mathematics and 
Theoretical Physics Grant No. JPMXP0619217849).

\appendix

%
%======================================%
%<<<<<<<<<<<< APPENDIX A  >>>>>>>>>>>>>>%
%======================================%
%
    \section{Falloff behavior of the Christoffel symbols}
    \label{appendixa}
    \noindent Here, we list the falloff properties of the Christoffel symbols. For $n\geq4$, the components of the Christoffel symbols are estimated as 
    \begin{eqnarray}
    \Gamma^{u}_{uu}%&=& \dot{B} -\frac{1}{2}A_{,r}-\frac{1}{2}AB_{,r}+\frac{1}{2}e^{-B}\left(h_{IJ}C^IC^J r^2\right)_{,r}\nonumber\\
     %\hspace{6.7mm}
     &=&\frac{n-2}{4}A^{(1)}r^{-n/2}+\left(\dot{B}^{(1)}-\frac{n-3}{2}m\right)r^{-(n-2)}+\Cr{-(n+1)/2}, \quad \nonumber\\
    \Gamma^{u}_{ur}&=&  0, \quad 
    \Gamma^{u}_{rr}=   0, \quad \nonumber\\
     \Gamma^{u}_{uI}&=& %\frac{1}{2}B_{,I}+\frac{1}{2}e^{-B}\left(h_{IJ}r^2C^J\right)_{,r} =
      -\frac{n-4}{4}C^{(1)}_{~~I}r^{-(n/2-1)}+ \Cr{-(n-1)/2}, \quad 
    \nonumber \\
    \Gamma^{u}_{rI}&=&   0, \quad 
     \Gamma^{u}_{IJ}%=\frac{1}{2}e^{-B}\left(h_{IJ}r^2\right)_{,r}
     \ =\ \omega_{IJ} r -\frac{n-6}{4}h^{(1)}_{IJ}r^{-(n/2-2)} +\Cr{-(n-3)/2}, \quad 
    \nonumber \\
    \Gamma^{r}_{uu}%&=& \frac{1}{2}Ae^{-B}\left[2\left(-e^{-B}\right)_{,u}-\left(-Ae^B+h_{IJ}C^IC^J r^2\right)_{,r}\right]-\frac{1}{2}e^{-B}\left(-Ae^B+h_{IJ}C^IC^J r^2\right)_{,u}
    %\nonumber \\
    %&&
    %\hspace{6mm}+\frac{1}{2}C^Ie^{-B}\left[2r^2\left(h_{IJ}C^J\right)_{,u}-\left(-Ae^B+r^2h_{JK}C^JC^K\right)_{,I}\right]
    %\nonumber \\
    &=& \frac{1}{2}\dot{A}^{(1)}r^{-(n/2-1)}-\frac{1}{2}\dot{m}r^{-(n-3)} + \Cr{- (n-1)/2}, \quad 
    \nonumber \\
    \Gamma^{r}_{ur}%&=&-\frac{1}{2}e^{-B}\left(-Ae^B+h_{IJ}C^IC^J r^2\right)_{,r}+\frac{1}{2}C^{I}e^{-B}\left[r^2\left(h_{IJ}C^J\right)_{,r}-e^BB_{,I}\right]\nonumber \\
    &=&- \frac{n-2}{4}A^{(1)}r^{-n/2}+\frac{n-3}{2}mr^{-(n-2)}+\Cr{-(n+1)/2},\nonumber\\
     \Gamma^{r}_{rr}&=& %B_{,r}=
      -(n-2)B^{(1)}r^{-(n-1)}+\Cr{-(n-1/2)}, \quad 
    \nonumber \\
    \Gamma^{r}_{uI}%&=& \frac{1}{2}A\left[ B_{,I}-e^{-B}\left(h_{IJ}r^2C^J\right)_{,r}\right]-\frac{1}{2}e^{-B}\left(-Ae^B+r^2h_{JK}C^JC^K
    %\right)_{,I} \nonumber \\
    %&&\hspace{6mm}+\frac{1}{2}C^Je^{-B}\left[r^2\dot{h}_{IJ}+r^2\left(h_{JK}C^K\right)_{,I}-r^2\left(h_{IK}C^K\right)_{,J}\right] \quad 
    %\nonumber \\
    &=& \left(\frac{n-4}{4}C^{(1)}_{~~I}+\frac{1}{2}A^{(1)}_{,I}\right)r^{-(n/2-1)}+\left(-\frac{1}{2}m_{,I}+\frac{1}{2}C^{(1)J}\dot{h}^{(1)}_{IJ}\right)r^{-(n-3)}  \nonumber \\
    &&\hspace{6mm}+\Cr{-(n-1)/2} ,\nonumber\\
     \Gamma^{r}_{rI} &=& %\frac{1}{2}B_{,I}-\frac{1}{2}e^{-B}h_{IJ}r^2C^J_{,r}=
      \frac{n}{4}C^{(1)}_{~~I}r^{-(n/2-1)}+\Cr{-(n-1)/2}, \quad 
    \nonumber \\
    \Gamma^{r}_{IJ}%&=& -\frac{1}{2}e^{-B}h_{JK}C^K_{,I}r^2-\frac{1}{2}e^{-B}h_{IK}C^K_{,J}r^2+\frac{1}{2}e^{-B}r^2\dot{h}_{IJ}-\frac{1}{2}Ae^{-B}\left(h_{IJ}r^2\right)_{,r}-\frac{1}{2}C^Ke^{-B}h_{IJ,K}r^2\nonumber\\
    &=& -\omega_{IJ} r+\frac{1}{2}\dot{h}^{(1)}_{IJ}r^{-(n/2-3)} +\Cr{-(n-5)/2}, \quad 
    \nonumber \\
    \Gamma^{I}_{uu}%&=& \frac{1}{2}C^Ie^{-B}\left[2e^BB_{,u}-\left(-Ae^B+r^2h_{JK}C^JC^K\right)_{,r}\right]\nonumber\\
    %&&\hspace{6mm}+\frac{1}{2}h^{IJ} r^{-2}\left[2\left(r^2h_{JK}C^K\right)_{,u}-\left(-Ae^B+r^2h_{KL}C^KC^L\right)_{,J}\right]
    %\nonumber\\
    &=&\dot{C}^{(1)I}r^{-n/2}+ \Cr{-(n+1)/2} ,\quad 
    \nonumber \\
     \Gamma^{I}_{ur}&=&%\frac{1}{2}h^{IJ} r^{-2}\left[\left(r^2h_{JK}C^K\right)_{,r}+e^B B_{,J}\right]=
      -\frac{n-4}{4}C^{(1)I}r^{-(n/2+1)}+ \Cr{-(n+3)/2}, \quad 
    \Gamma^{I}_{rr}=   0, \quad 
    \nonumber \\
    \Gamma^{I}_{uJ}%&=& -\frac{1}{2}C^Ie^{-B}\left[e^BB_{,J}+\left(r^2h_{JK}C^K\right)_{,r}\right]+\frac{1}{2}h^{IK}\left[\left(h_{KL}C^L\right)_{,J}+\dot{h}_{KJ}-\left(h_{JL}C^L\right)_{,K}\right] 
    %\nonumber \\
    &=& \frac{1}{2}\dot{h}^{(1)I}_{~~~J}r^{-(n/2-1)}+\Cr{-(n-1)/2},\nonumber\\
     \Gamma^{I}_{rJ}&=& %\frac{1}{2}r^{-2}h^{IK}\left(r^2h_{JK}\right)_{,r}
      %\nonumber\\
      %\ =\
      \delta^I_Jr^{-1} -\frac{n-2}{4}h^{(1)I}_{~~~J}r^{-n/2} +\Cr{-(n+1)/2},  \nonumber\\
  %  \Gamma^{I}_{JK}%&=&-\frac{1}{2}C^Ie^{-B}\left(r^2h_{JK}\right)_{,r}+\frac{1}{2}h^{IL}\left(h_{JL,K}+h_{KL,J}-h_{JK,L}\right)
  %  %\nonumber\\
  %  &=&
  %  \frac{1}{2}\omega^{IL}\left(\omega_{JL,K}+\omega_{KL,J}-\omega_{JK,L}\right)-\left[C^{(1)I}\omega_{JK}+\frac{1}{2}h^{IL(1)}\left(\omega_{JL,K}+\omega_{KL,J}-\omega_{JK,L}\right)\right.\nonumber\\
  %  &&\left.\hspace{6mm}-\frac{1}{2}\omega^{IL}\left(h^{(1)}_{JL,K}+h^{(1)}_{KL,J}-h^{(1)}_{JK,L}\right)\right]r^{-(n/2-1)}+\Cr{-(n-1)/2},\label{conhd}
   \Gamma^{I}_{JK}%&=&-\frac{1}{2}C^Ie^{-B}\left(r^2h_{JK}\right)_{,r}+\frac{1}{2}h^{IL}\left(h_{JL,K}+h_{KL,J}-h_{JK,L}\right)
    %\nonumber\\
    &=&{}^{(\omega)}\Gamma^I_{JK}
    -\left[C^{(1)I}\omega_{JK}-\frac{1}{2}\left(D_Jh^{(1)I}_{~~~K}+D_Kh^{(1)I}_{~~~J}-D^Ih^{(1)}_{~JK}\right)\right]r^{-(n/2-1)}\nonumber \\
    &&\hspace{6mm}+\Cr{-(n-1)/2},\label{conhd}
    \end{eqnarray}
    where we used $n\geq4$; ${}^{(\omega)}\Gamma^I_{JK}$ is the Christoffel symbol with respect to $\omega_{IJ}$, that is, ${}^{(\omega)}\Gamma^I_{JK}:=\frac{1}{2}\omega^{IL}\left(\omega_{JL,K}+\omega_{KL,J}-\omega_{JK,L}\right)$; and $D_I$ is the covariant derivative with respect to $\omega_{IJ}$.

    In particular, in four dimensions, the components of the Christoffel symbols are written as follows:
    \begin{eqnarray}
    \Gamma^{u}_{uu} &=& \left(\dot{B}^{(1)} -\frac{1}{2}m\right)r^{-2}+\Cr{-3}   , \quad 
    \Gamma^{u}_{ur}=  0, \quad 
    \Gamma^{u}_{rr}=   0, \quad 
    \Gamma^{u}_{uI}=   \Cr{-2}, \quad 
    \nonumber \\
    \Gamma^{u}_{rI}&=&   0, \quad 
    \Gamma^{u}_{IJ}= \omega_{IJ} r +\frac{1}{2}h^{(1)}_{IJ} +\Cr{-1}, \quad 
    \Gamma^{r}_{uu}= - \frac{1}{2}\dot{m}r^{-1} +  \Cr{-2}, \quad 
    \nonumber \\
    \Gamma^{r}_{ur}&=&   \frac{1}{2}mr^{-2}+\Cr{-3}, \quad 
    \Gamma^{r}_{rr}=   -2B^{(1)}r^{-3}+\Cr{-4}, \quad \nonumber\\
    \Gamma^{r}_{uI}&=&  \left(-\frac{1}{2}m_{,I}+\frac{1}{2}C^{(1)J}\dot{h}^{(1)}_{IJ}\right)r^{-1}+\Cr{-2}, \quad 
    \Gamma^{r}_{rI}= C^{(1)}_{~~I}r^{-1}+\Cr{-2}, \quad 
    \quad 
    \nonumber\\
    \Gamma^{r}_{IJ}&=& -\left(\omega_{IJ}  -  \frac12 \dot{h}^{(1)}_{IJ} \right)r + \Cr{0}\label{Chfall4d}, \quad 
    \Gamma^{I}_{uu}=   \dot{C}^{(1)I}r^{-2}+ \Cr{-3}, \quad 
    \Gamma^{I}_{ur}=  \Cr{-4}, \quad 
    \Gamma^{I}_{rr}=   0, \quad 
    \nonumber\\
    \Gamma^{I}_{uJ}&=&   \frac{1}{2}\dot{h}^{(1)I}_{~~~J}r^{-1}+\Cr{-2}, \quad 
     \Gamma^{I}_{rJ}=  \delta^I_Jr^{-1} -\frac12h^{(1)I}_{~~~J}r^{-2}+\Cr{-3},  \quad
    \nonumber\\
    \Gamma^{I}_{JK}&=&{}^{(\omega)}\Gamma^{I}_{JK} -\left[C^{(1)I}\omega_{JK}
      -\frac{1}{2}\left(D_Jh^{(1)I}_{~~~K}+D_Kh^{(1)I}_{~~~J}-D^Ih^{(1)}_{~JK}\right) \right]r^{-1}+\Cr{-2}. 
    \end{eqnarray}

%
%======================================%
%<<<<<<<<<<<< APPENDIX B  >>>>>>>>>>>>>>%
%======================================%
%
    
    \section{Upper bounds of $r'$ and $\left(x^I\right)^\prime $}
    \label{appendixb}
    In this Appendix, we will derive 
     the upper bounds of $r'$ and $\left(x^I\right)^\prime $ for $n\geq4$.
Using $u'=\Cr{-2}$, we give the upper bound of $r'$.
From the null condition~\eqref{null4d} and the inequality of Eq.~\eqref{u'est}, which are valid for $n\geq4$, $\left|\left(x^I\right)^\prime \right|$ can be estimated as 
\begin{eqnarray}
    0\le \left|\left(x^I\right)^\prime \right|^2  &=& \left| \red{\left(x^I\right)^\prime} + C^I {u'} - C^I {u'} \right|^2 \le 
    2\left| \left(x^I\right)^\prime  + C^I {u'}\right|^2+ 2\left| C^I {u'} \right|^2 
    \nonumber \\
    &&\hspace{3mm}=  \Cr{-2}{u'}^2 + \Cr{-2} u' r'  + 2|C^I|^2 {u'}^2= \Cr{-6}+ \Cr{-4}r'.
    \label{ineqxI}
    \end{eqnarray} 
    Then, the geodesic equation of $r$ for $n\geq4$ is\footnote{\red{For $n=4$, from the difinition, $A^{(1)}$ is set to be zero because of its absence.}}
\begin{eqnarray}
    r'' &=&  \left[-\frac{1}{2}\dot{A}^{(1)}r^{-(n/2-1)}+\frac{1}{2}\dot{m}r^{-(n-3)} + \Cr{- (n-1)/2}\right] {u^\prime}^2\nonumber\\
    &&\hspace{12mm}+\left[\frac{n-2}{2}A^{(1)}r^{-n/2}-\left(n-3\right)mr^{-(n-2)}+\Cr{-(n+1)/2}\right] u'r'\nonumber\\
    &&\hspace{12mm}+\left[(n-2)B^{(1)}r^{-(n-1)}+\Cr{-(n-1/2)}\right] {r^\prime}^2\nonumber\\
    &&\hspace{12mm}+\bigg[\left(-\frac{n-4}{2}C^{(1)}_{~~I}-A^{(1)}_{,I}\right)r^{-(n/2-1)}+\left(m_{,I}-C^{(1)J}\dot{h}^{(1)}_{IJ}\right)r^{-(n-3)}  \nonumber \\
    &&\hspace{20mm}+\Cr{-(n-1)/2}\bigg] u' \left(x^I\right)^\prime  
    -\left[\frac{n}{2}C^{(1)}_{~~I}r^{-(n/2-1)}+\Cr{-(n-1)/2}\right] r' \left(x^I\right)^\prime  \nonumber\\
    &&\hspace{12mm}+
    \left[\omega_{IJ} r-\frac{1}{2}\dot{h}^{(1)}_{IJ}r^{-(n/2-3)} +\Cr{-(n-5)/2} \right] \left(x^I\right)^\prime \left(x^J\right)^\prime \nonumber\\
    &=& \left[-\frac{1}{2}\dot{A}^{(1)}r^{-(n/2-1)}+\frac{1}{2}\dot{m}r^{-(n-3)} + \Cr{- (n-1)/2}\right] {u^\prime}^2\nonumber\\
 &&\hspace{12mm}+\left[\frac{n-2}{2}A^{(1)}r^{-n/2}\red{-}(n-3)mr^{-(n-2)}+\Cr{-(n+1)/2}\right] u'r'
  \nonumber\\
  &&\hspace{12mm}
+\Cr{-(n-1)}{r^\prime}^2+\Cr{1}\left|\left(x^I\right)^\prime \right|^2\nonumber\\
 &=& \Cr{-(n/2+3)}+\Cr{-(n+4)/2}r^\prime+\Cr{-(n-1)}{r^\prime}^2
  +\Cr{-5}+ \Cr{-3}r'\nonumber\\
    &=&\Cr{-3}+ \Cr{-3}{r^\prime}^2\label{r''<},
\end{eqnarray}
where we used Eqs.~\eqref{conhd} in the first equality; the arithmetic-geometric mean inequality 
\begin{eqnarray}
&&\Cr{-(n/2-1)}u' \left|\left(x^I\right)^\prime \right| \le \frac{1}{2}\left[\Cr{-(n-2)}{u'}^2 + \left|\left(x^I\right)^\prime \right|^2\right]
\end{eqnarray}
and
\begin{eqnarray}
 \Cr{-(n/2-1)}r' \left|\left(x^I\right)^\prime \right| &=& \Cr{-(n-1)/2}r' \Cr{1/2} \left|\left(x^I\right)^\prime \right|\nonumber\\
&& \le \frac{1}{2}\left[\Cr{-(n-1)}{r'}^2 +\Cr{1} \left|\left(x^I\right)^\prime \right|^2\right]
\end{eqnarray}
in the second equality, and Eqs.~\eqref{u'est},~\eqref{u'esthd} and~\eqref{ineqxI} in the third equality; and  
\begin{eqnarray}
 \Cr{-3} r' &=& \Cr{-3/2}\Cr{-3/2}r'\nonumber\\
 &&\le \frac{1}{2}\left[\Cr{-3} + \Cr{-3}{r^\prime}^2\right]
\end{eqnarray}
in the fourth equality. 
Equation~\eqref{r''<} means 
\begin{eqnarray}
r''< {\bar C_1} \frac{{r'}^2 + {\bar C_2}}{2r'}  \frac{r'}{r^3},
\end{eqnarray}
with positive constants ${\bar C_1}$ and ${\bar C_2}$.
This gives
\begin{eqnarray}
\log \left({r'}^2 + {\bar C_2}\right)'< {\bar C_1}   \frac{r'}{r^3}.
\end{eqnarray}
Integrating this, we have 
\begin{eqnarray}
\log \left({r'}^2 + {\bar C_2}\right)< {\bar C_3} - {\bar C_1}   \frac{1}{2r^2},
\end{eqnarray}
where ${\bar C_3}$ is the integration constant. 
From this inequality, we have
\begin{eqnarray}
{r'}^2 + {\bar C_2}<  \exp \left({\bar C_3} - {\bar C_1}   \frac{1}{2r^2} \right) <  \exp {\bar C_3},
\end{eqnarray}
that is 
\begin{eqnarray}
{r'}^2  <  \left(\exp {\bar C_3} \right) - {\bar C_2}=: {\bar C_4}.
\end{eqnarray}
Here ${\bar C_4}$ should be positive.
This means $r'$ has a positive upper bound
\begin{eqnarray}
    \label{r'<}
{r'}  <  \sqrt{{\bar C_4}}. \label{r'ineq}
\end{eqnarray}
Integration of this gives
\begin{eqnarray}
    \label{r<}
r < \sqrt{{\bar C_4}} \lambda + {\bar C_5},
\end{eqnarray}
where ${\bar C_5}$ is the integration constant. 

In addition, inequalities of Eqs.~\eqref{ineqxI} and~\eqref{r'ineq} give
\begin{eqnarray}
&&0\le \left|\left(x^I\right)^\prime \right|^2 = \Cr{-6}+ \Cr{-4}r' = \Cr{-4}.
\end{eqnarray}
Therefore, we have 
\begin{eqnarray}
    \label{xI<}
&&\left(x^I\right)^\prime = \Cr{-2}.
\end{eqnarray}

\end{document}